\PassOptionsToPackage{dvipsnames}{xcolor}
\let\EEGJevPublisherBibStyle\bibliographystyle
\def\bibliographystyle#1{\EEGJevPublisherBibStyle{eegjev-nature}}
\documentclass[pdflatex,sn-nature]{sn-jnl}
\unnumbered
\let\bibliographystyle\EEGJevPublisherBibStyle
\usepackage[T1]{fontenc}
\usepackage[utf8]{inputenc}
\usepackage{microtype,graphicx,booktabs}
\usepackage{float}
\usepackage{eegjev-review}
\usepackage{amsmath,amssymb,mathtools}
\usepackage{tabularx,array,xcolor}
\usepackage{eegjev-tables}
\setcitestyle{super,open={},close={}}
\usepackage{longtable,placeins,bookmark}
\newcommand{\suppref}[1]{\ref{#1}}
\newcommand{\EEGJevResourceLink}[3]{%
  \href{#1}{\raisebox{-0.35ex}{\includegraphics[height=2.2ex]{figures/resource-icons/#2.pdf}}\enspace #3}}
\newcommand{\EEGJevAuthorLinks}{%
  \EEGJevResourceLink{https://github.com/dengzhe-hou/EEGJev}{github}{GitHub}%
  \hspace{1.2em}\EEGJevResourceLink{https://dengzhe-hou.github.io/EEGJev/}{home}{Homepage}%
  \hspace{1.2em}\mbox{%
    \EEGJevResourceLink{https://huggingface.co/dengzhe-hou/EEGJev}{huggingface}{Hugging Face Model}%
    \enspace\textperiodcentered\enspace
    \href{https://huggingface.co/datasets/dengzhe-hou/EEGJev-Bench}{Bench}}}
\makeatletter
\patchcmd{\@maketitle}{\artauthors\par}
  {\artauthors\par\vskip4pt{\normalfont\small\EEGJevAuthorLinks\par}}
  {}{\PackageError{eegjev-review}{Could not place the project links beneath the author line}{}}
\patchcmd{\@maketitle}{\global\punctcount\aucount\vskip20pt}
  {\global\punctcount\aucount\vskip10pt}
  {}{\PackageError{eegjev-review}{Could not adjust the gap below the title}{}}
\patchcmd{\@maketitle}{\vskip24pt}{\vskip0pt}
  {}{\PackageError{eegjev-review}{Could not adjust the first affiliation gap}{}}
\patchcmd{\@maketitle}{\vskip24pt}{\vskip6pt}
  {}{\PackageError{eegjev-review}{Could not adjust the abstract gap}{}}
\patchcmd{\@maketitle}{\vskip36pt}{\vskip6pt}
  {}{\PackageError{eegjev-review}{Could not adjust the title block gap}{}}
\makeatother
\begin{document}
\title{EEGJev: A System One Decision Model and Benchmark for Multi-Task EEG Question Answering}
\author*[1,2]{\fnm{Dengzhe} \sur{Hou}}
\author[1,2]{\fnm{Kazunori D.} \sur{Yamada}}
\affil[1]{\orgdiv{Graduate School of Information Sciences}, \orgname{Tohoku~University},~\orgaddress{\city{Sendai},~\country{Japan}}}
\affil[2]{\orgdiv{Unprecedented-scale Data Analytics Center}, \orgname{Tohoku~University},~\orgaddress{\city{Sendai},~\country{Japan}}}
\abstract{A single electroencephalographic (EEG) recording can support multiple decoding tasks.
Fixed-label tests cannot show how EEG foundation models handle changing questions or options.
We introduce \textbf{EEGJev}, a fast decision model that scores candidate answers in one pass.
Our benchmark, \textbf{EEGJev-Bench}, covers six question types across 19 public EEG sources.
We held the recording and options fixed while switching between questions about attributes paired during training, such as facial expression and the hand participants were asked to use.
Across 13 test sources, EEGJev selected a label for the requested attribute with 99.70\% accuracy (49.20\% under uniform choice).
Balanced accuracy for the specific answer was 62.04\%.
Its 95th-percentile response time was shorter than NeuroLM's on matched validation questions answered individually.
EEGJev lets users switch decoding targets by changing the question.
With EEGJev-Bench, researchers can test whether new decoders follow the question more reliably, identify the correct answer more often, or both.}
\newcommand{\EEGJevKeywords}{EEG, benchmark, question answering, refusal, foundation models}
\keywords{\EEGJevKeywords}
\maketitle
\vspace{4pt}
\noindent We introduce EEGJev, a question-guided EEG decision model (Figure~\ref{fig:model}).\par
\begingroup
\setlength{\intextsep}{4pt}
\begin{figure}[H]
\centering
\renewcommand{\figurecaptionfont}{\normalfont\fontsize{9.5}{11.25}\selectfont}
\includegraphics[width=\textwidth]{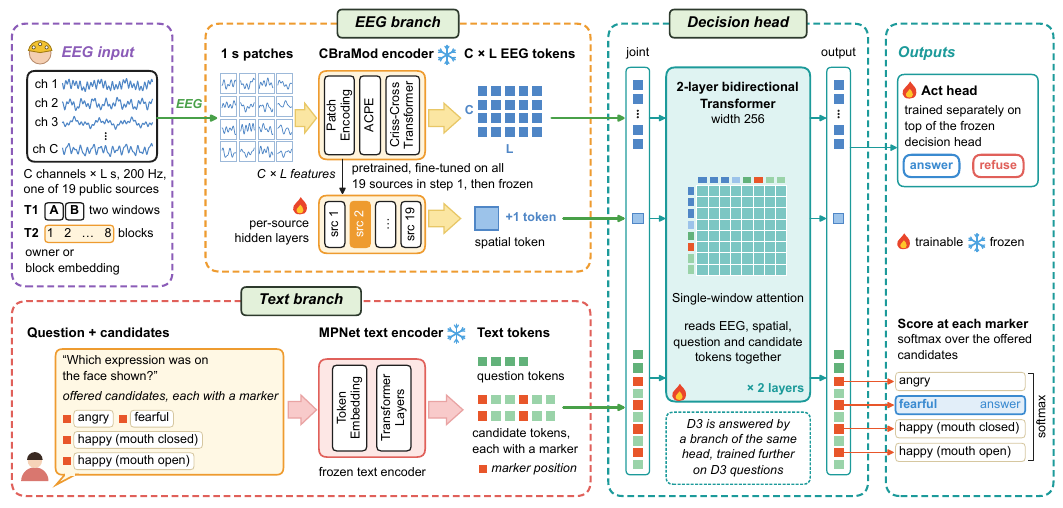}
\par\vspace{0pt}
\caption{EEGJev. CBraMod, fine-tuned on all sources in step 1 and then frozen, turns the window into $C \times L$
tokens, and per-source hidden layers add one spatial token. Frozen MPNet reads the question and the offered candidates,
each candidate with a marker position. A two-layer bidirectional transformer reads all tokens together and scores each
candidate at its marker, with a softmax over the offered candidates. A separately trained Act head decides whether to
answer, and D3 is answered by a branch of the same head trained further on D3 questions. For T1 the head reads the
tokens of two windows and for T2 those of up to eight ordered blocks of at most 4\,s, each marked by an owner or
block embedding.
Direct attention from candidate tokens to EEG tokens is restricted to the candidate's window in T1 and to blocks overlapping its half in T2.}
\label{fig:model}
\end{figure}
\endgroup
\clearpage

Electroencephalographic (EEG) decoding tests which aspects of perception and action can be inferred from scalp recordings.
An experimental trial can link a recording to several annotations, including the stimulus presented and the response that the participant was asked to make.
For the same second of EEG in Figure~\ref{fig:bench}, a decoder should answer ``fearful'' when asked about the facial expression and ``left hand'' when asked about the instructed response.
Selecting a task-specific classifier fixes the prediction target in advance.
A researcher who wants to predict a different attribute of the same trial must then select another classifier or output head.
A decoder that accepts a question would let the researcher specify the target when analyzing the recording.
Can one decoder follow a change in the requested attribute for the same recording, and how often does it choose the correct answer?

\begin{samepage}
LaBraM~\cite{labram2024} and CBraMod~\cite{wang2025cbramod} reuse pretrained EEG encoders across tasks, usually with a separate head for each prediction target.
NeuroLM~\cite{neurolm2024} generates answers from instructions.
LEAF~\cite{leaf2025} uses the instruction to score candidate labels, and METIS~\cite{metis2026} scores the offered options.
Evaluating such models with a different choice set for each question can make the target apparent from the options.
\par
\end{samepage}

\begin{samepage}
For example, choices that contain only expressions or only hand labels reveal which attribute is being asked about.
Keeping the EEG and the ordered list of answer options fixed lets us test whether a change in the question changes the attribute of the selected answer.
There is also a distinction between asking two questions and requiring two decisions.
PhysioBench~\cite{physiobench2026} asks multiple questions about one EEG window.
For its paired seizure questions, identifying the seizure state also answers whether a seizure is present
(\href{https://github.com/Leanna97/PhysioBench/blob/e512592c6cf8c413e4f7d579b14316f2f8a684aa/scripts/data/builders/EEG/EEG_CHBMIT/build_manifest.py}{released construction code}).
BrainBench~\cite{brainbench2026} includes refusal cases without matched answerable controls.
Pairing a refusal item with an answerable version tests whether a decoder also responds when a correct option is available.
\par
\end{samepage}

\begin{samepage}
A shared CBraMod encoder represents the signal, and a frozen MPNet encoder~\cite{song2020mpnet} represents the question and candidate answers.
Jev and Laya are System One models that return decisions directly~\cite{hou2026s1cogbench}.
Following Laya's design~\cite{laya_code}, EEGJev's decision head reads the EEG and text representations together and scores every offered answer in one pass.
Changing the question can thus query a different attribute through the same scoring interface.
A separately trained head decides whether to answer, and a branch of the decision head receives further training for ordered categories such as age groups.
For two-window comparisons and long-window questions, the decision head uses information about which window or time block the signal tokens come from.
The same head thus scores window and time choices as well as attribute labels.
\par
\end{samepage}

We also introduce EEGJev-Bench, which covers six question types across 19 public EEG sources.
To build it, we checked the public releases against their original datasets and applied a common processing pipeline to obtain EEG windows.
Within each scored source, different participants provide the training and test windows.
To assess two predictions from one window, we pair questions whose answers do not determine one another in that source's training recordings.
The pair score requires a correct answer to each question, measuring whether the model can answer both on the same trial.
To test refusal, we remove the correct option from one version and a distractor from the other.
The two versions share the EEG and the question and offer the same number of choices, with the correct option available in one version.
The other tasks ask about a single attribute or an ordered category, compare two windows, or locate an event within a longer window.\par
\clearpage
\begingroup
\setlength{\intextsep}{4pt}
\begin{figure}[H]
\centering
\renewcommand{\figurecaptionfont}{\normalfont\fontsize{9.5}{11.25}\selectfont}
\includegraphics[width=\textwidth]{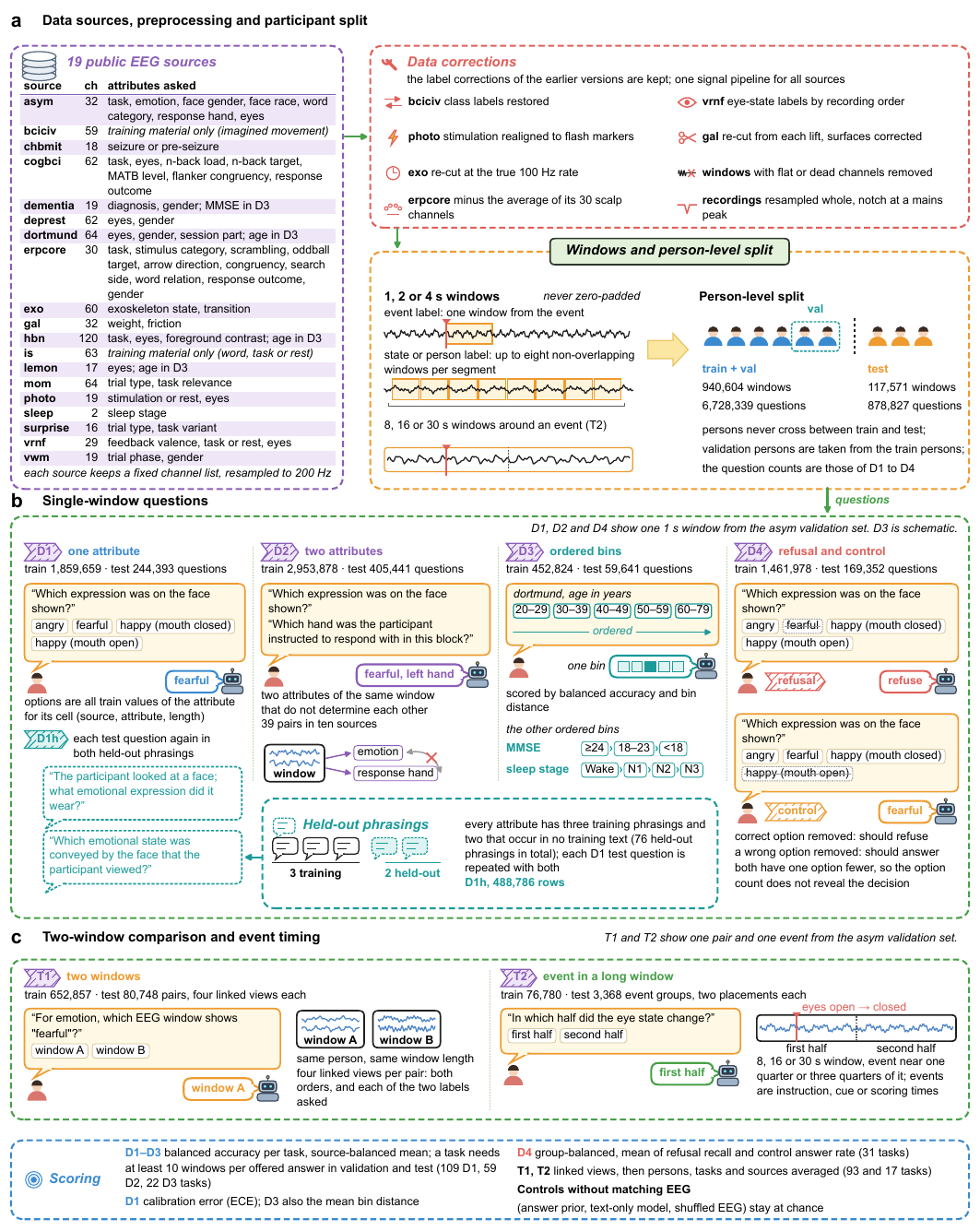}
\par\vspace{0pt}
\caption[EEGJev-Bench.]{\textbf{EEGJev-Bench.} \textbf{a}, We checked the public EEG releases against their original datasets and corrected
discrepancies when building the benchmark. Within each scored source, training and test windows come
from different participants. Each question pairs one or two EEG windows with a prompt and answer
choices. D1--D4 and T1 use windows of 1, 2 or 4\,s.
\textbf{b}, D1 asks about one attribute. D2 asks two questions about the
same window, chosen so that knowing one answer does not reveal the other. D3 asks the model to select an
ordered category, such as an age group. D4 tests refusal by removing the correct option. A matched,
answerable version removes one incorrect option. D1h repeats each D1 test question using two phrasings absent from training.
\textbf{c}, T1 asks which of two windows shows a given label. T2 asks whether an event occurs in the first or second half of an 8, 16 or 30\,s window.
The D3 example uses the Dortmund age bins. The highlighted bin illustrates one possible answer. The
other five examples come from the \emph{asym} validation set.}
\label{fig:bench}
\end{figure}
\endgroup
\raggedbottom
For each task type, the aggregate gives each scored source equal weight, regardless of how many windows it contains.

\clearpage
\setlength{\textfloatsep}{9pt plus 2pt minus 1pt}
\setlength{\floatsep}{9pt plus 2pt minus 1pt}
\setlength{\intextsep}{7pt plus 2pt minus 1pt}

We changed the question while keeping the EEG window and answer options fixed.
For attribute pairs seen during training, EEGJev almost always chose a label for the requested attribute.
Choosing the correct answer was harder.
On the same binary validation questions answered individually, its 95th-percentile response time was shorter than NeuroLM-B's on both a single CPU core and a GPU.
EEGJev lets users choose what a single decoder should identify in a recording by changing the question.
EEGJev-Bench gives researchers a common test of whether a new decoder follows the question more reliably, chooses the correct answer more often, or both.

\section{Results}
\label{sec:results}

\subsection{Question-guided selection of familiar attributes}
\label{sec:res-controls}

We first held both the EEG window and the ordered list of options fixed and changed only which attribute the question asked about. Each list contained four to ten options from two disjoint label sets. All tested attribute pairs appeared in training.
Across 13 sources, EEGJev chose a label for the requested attribute with 99.70\% test accuracy (49.20\% for uniform choice).
Balanced accuracy for the specific answer was 62.04\% (Figure~\ref{fig:controls-transfer}a). When both questions had to be answered correctly, the source-balanced score was 44.37\%.

We also tested how EEGJev responds when the question points to the wrong attribute. On D1, we replaced the question with one about another task
from the same source, keeping the EEG window, candidates and reference answer fixed (Figure~\ref{fig:controls-transfer}b).
Test balanced accuracy falls by 12.6 percentage points (95\% interval for the change, $-14.1$ to $-11.1$).
An empty question lowers the score by 1.3 points, an untrained task code by 1.9 and the two held-out phrasings
by 2.9 and 1.3. Scores are stable under candidate rotation. With letter codes in place of candidate names, test
balanced accuracy is 43.5\% with or without the mapping in the question, close to the controls without matching
EEG. Training used the mapped format for 10\% of D1 and D2 questions. Validation shows the same pattern. The question guides attribute selection, and candidate names help
EEGJev identify the specific answer.

\subsection{Answer accuracy on familiar tasks}
\label{sec:res-d1}

On the D1 test tasks, balanced accuracy averaged across sources was 63.8\% for EEGJev and 64.5\% for
task-specific EEGNet (Table~\ref{tab:d1}). The paired difference was $-0.69$ percentage points (95\% interval
$-1.77$ to $+0.26$ from resampling the 217 test participants within sources). On validation, the corresponding
difference was $-2.0$ percentage points (95\% interval $-3.2$ to $-0.8$ from resampling tasks within sources).
The ordinary classifier on the same encoder reached 64.2\% on test, 0.3 points above EEGJev ($-0.5$ to $+1.1$).
LaBraM reached 64.3\% on the 96 tasks it covered and was 0.5 points below EEGJev on those tasks ($-1.5$ to $+0.5$).
After temperature scaling, EEGJev's D1 calibration error was 2.0\% (test) and 2.4\% (validation). LaBraM had
the highest test error (9.7\%).

\begingroup
\setlength{\intextsep}{5pt}
\begin{figure}[!t]
\centering
\includegraphics[width=\textwidth,trim=0 2.5mm 0 2.2mm,clip]{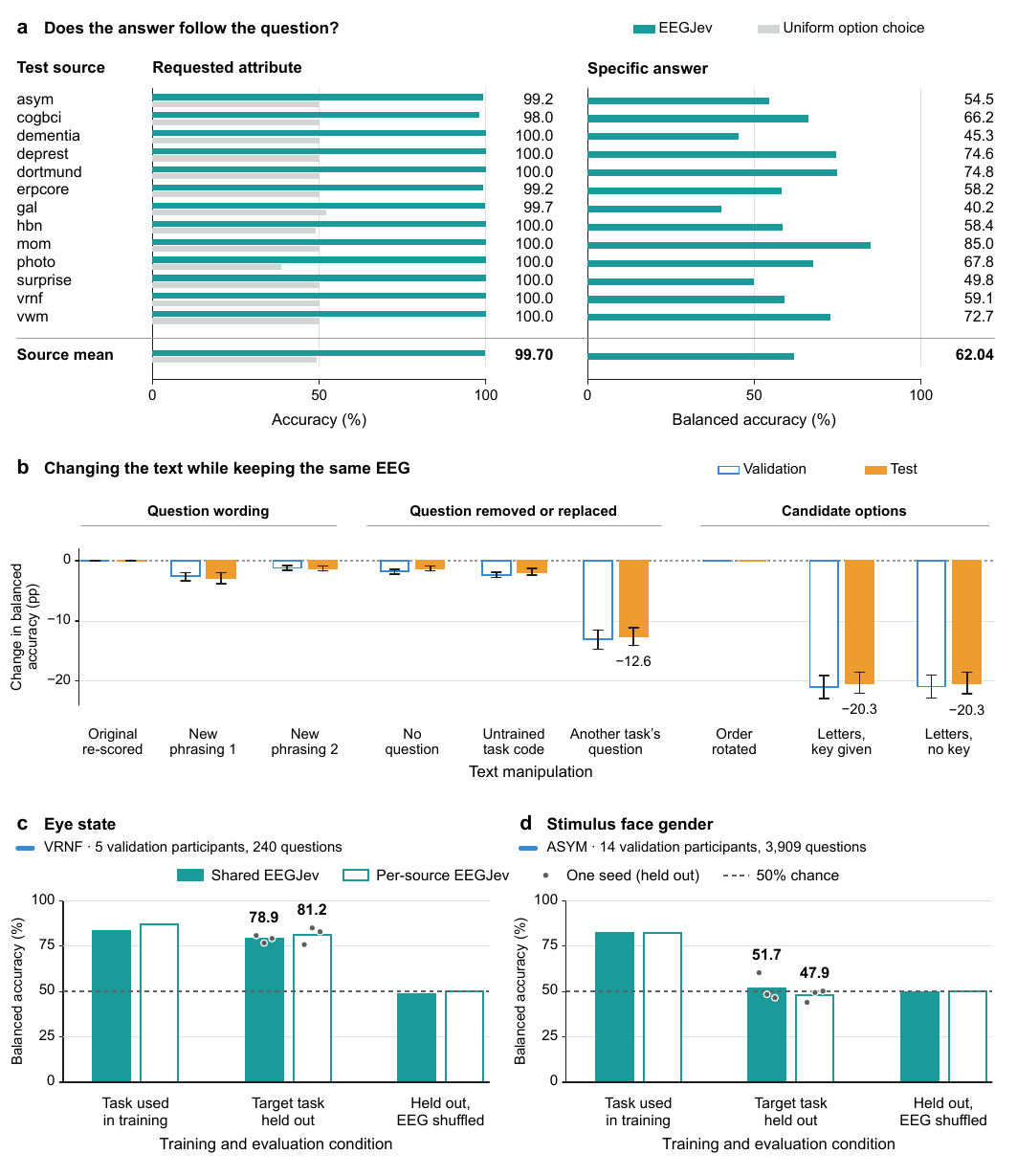}
\par\vspace{0pt}
\caption[Question following, text controls and transfer to held-out tasks.]{\textbf{Question following, text controls and transfer to held-out tasks.}\par
\textbf{a}, Two questions about familiar attributes share an EEG window and an ordered list of options. The left column counts a label as correct when it belongs to the requested attribute. EEGJev scores average three training seeds for each of 13 test sources (380,570 eligible questions per seed). The last row weights sources equally.\par
\textbf{b}, Changes are relative to scores with the original D1 questions and options. Whiskers show paired 95\% intervals from resampling tasks within sources. Letter codes replace the candidate text. The test set covers 109 tasks from 17 sources.\par
\textbf{c,d}, Bars show means across three training seeds.}
\label{fig:controls-transfer}
\end{figure}
\endgroup

Figure~\ref{fig:bench-model-overview} compares the six question types. NeuroLM was instruction-tuned on benchmark D1--D3 questions.
On the 99 D1 tasks it covers, NeuroLM scores 58.6\%, 5.4 points below EEGJev ($-6.6$ to $-4.3$).
With released weights, METIS and LEAF score 43.5\% to 45.0\%, close to controls without EEG (43.5\% to 43.8\%).

\begingroup
\setlength{\intextsep}{5pt}
\begin{figure}[!t]
\centering
\includegraphics[width=\textwidth]{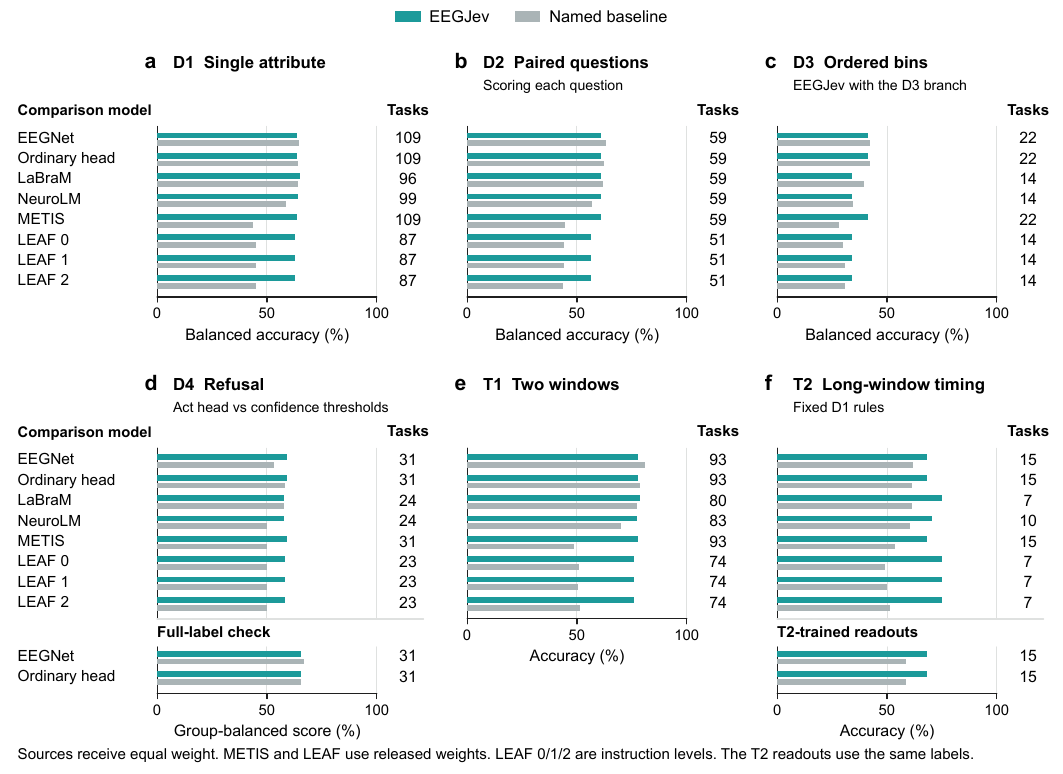}
\par\vspace{0pt}
\caption[Model comparisons across EEGJev-Bench.]{\textbf{Model comparisons across EEGJev-Bench.}
Each pair of bars compares EEGJev with the named baseline on the test tasks that both cover. The Tasks column gives this count. Scores weight sources equally.
\textbf{d}, The upper rows compare EEGJev's trained refusal head with confidence thresholds. The lower rows apply the full-label check to both models.
\textbf{e,f}, Fixed rules combine D1 outputs. The lower rows in \textbf{f} use time-ordered readouts trained on the same T2 labels as EEGJev.}
\label{fig:bench-model-overview}
\end{figure}
\endgroup

For D1h, we asked familiar D1 questions using wording held out from training.
With a shared encoder, EEGJev's test balanced accuracy falls from 63.8\% to 61.7\% (59.7\% to 62.9\% across seeds).
With a separate encoder for each source, it falls from 64.1\% to 59.8\% (53.7\% to 63.4\% across seeds). NeuroLM loses 3.1 points. The ordinary classifier
does not read the question and is unchanged (Supplementary Note~\suppref{app:d1h}, Supplementary Table~\suppref{tab:d1h}).

\begin{table*}[!htbp]
\centering
\caption{D1 source-balanced mean balanced accuracy (\%), seed means, each model over the tasks it covers. Differences
to EEGNet are paired over the Tasks column on common rows, with 95\% intervals from a bootstrap over tasks within
sources. LaBraM has no models for chbmit, hbn and sleep, NeuroLM no input for hbn and sleep, and LEAF no predictions
for chbmit, deprest, hbn, mom and sleep. The averaged row uses the rule described in
``\nameref{sec:res-d2d3d4}''. Validation is used for model selection where applicable.
Shading shows BA on all 109 tasks. Bold marks the highest BA within each split.}
\label{tab:d1}
\setlength{\tabcolsep}{3pt}
\EEGJevTableFont
\begin{tabular}{@{}J{121pt}rrrrr@{}}
\toprule
 & & \multicolumn{2}{c}{Test} & \multicolumn{2}{c}{Validation} \\
\cmidrule(lr){3-4}\cmidrule(l){5-6}
\EEGJevTableHead \textbf{Model} & \textbf{Tasks} & \textbf{BA} & \textbf{minus EEGNet} & \textbf{BA} & \textbf{minus EEGNet} \\
\midrule
\EEGJevTableGroup{6}{Models trained on EEGJev-Bench}
EEGNet, one model per task & 109 & \EEGJevTableScore{25}{64.5} &  & \EEGJevTableScore{25}{66.4} &  \\
EEGJev & 109 & \EEGJevTableScore{25}{63.8} & $-0.7$ ($-2.0$, $+0.6$) & \EEGJevTableScore{25}{64.4} & $-2.0$ ($-3.2$, $-0.8$) \\
Ordinary classifier & 109 & \EEGJevTableScore{25}{64.2} & $-0.4$ ($-1.5$, $+0.7$) & \EEGJevTableScore{25}{64.9} & $-1.5$ ($-2.5$, $-0.5$) \\
EEGJev and ordinary, averaged & 109 & \EEGJevTableScore{25}{\textbf{65.2}} & $+0.7$ ($-0.5$, $+1.9$) & \EEGJevTableScore{25}{65.9} & $-0.5$ ($-1.5$, $+0.6$) \\
LaBraM, one model per task & 96 & 64.3 & $-1.4$ ($-2.6$, $-0.1$) & 65.7 & $-2.6$ ($-3.7$, $-1.6$) \\
NeuroLM, instruction-tuned & 99 & 58.6 & $-6.5$ ($-8.2$, $-5.0$) & 58.9 & $-8.8$ ($-10.5$, $-7.2$) \\
\EEGJevTableGroup{6}{Models with fixed weights}
METIS, fixed weights & 109 & \EEGJevTableScore{19}{43.5} & $-21.0$ ($-23.3$, $-18.6$) & \EEGJevTableScore{18}{43.1} & $-23.3$ ($-25.3$, $-21.2$) \\
LEAF, level 0 & 87 & 44.8 & $-18.7$ ($-21.2$, $-16.2$) & 44.8 & $-21.4$ ($-23.6$, $-19.1$) \\
LEAF, level 1 & 87 & 45.0 & $-18.6$ ($-21.1$, $-16.2$) & 45.3 & $-20.9$ ($-23.2$, $-18.5$) \\
LEAF, level 2 & 87 & 44.8 & $-18.7$ ($-21.2$, $-16.3$) & 45.5 & $-20.7$ ($-23.1$, $-18.4$) \\
\midrule
\EEGJevTableGroup{6}{Per-source encoders}
EEGJev, per-source encoders & 109 & \EEGJevTableScore{25}{64.1} & $-0.5$ ($-1.5$, $+0.6$) & \EEGJevTableScore{25}{65.3} & $-1.2$ ($-2.1$, $-0.2$) \\
Ordinary, per-source encoders & 109 & \EEGJevTableScore{25}{64.7} & $+0.2$ ($-0.7$, $+1.2$) & \EEGJevTableScore{25}{\textbf{66.9}} & $+0.5$ ($-0.8$, $+1.7$) \\
\bottomrule
\end{tabular}
\end{table*}

\begin{figure}[!htbp]
\centering
\includegraphics[width=\textwidth]{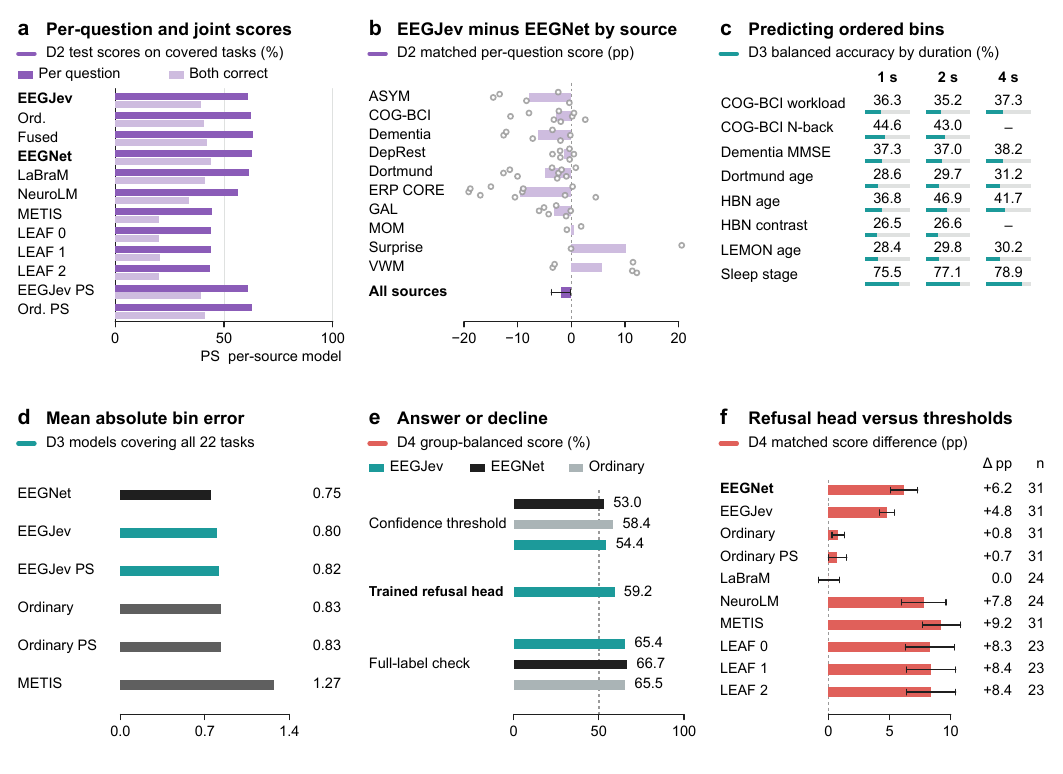}
\par\vspace{0pt}
\caption[Paired questions, ordered bins and refusal.]{\textbf{Paired questions, ordered bins and refusal.}
\textbf{a}, Dark and light bars show balanced accuracy per question and joint accuracy, respectively. The joint score balances across observed answer combinations. Each model is scored on the tasks it covers. Ordinary denotes the ordinary classification head, and fused denotes its equal-weight combination with EEGJev. PS denotes a per-source model.
\textbf{b}, Each open circle is one of 59 matched tasks. Bars show source means and the equally weighted mean over ten sources.
\textbf{c}, Colour bars and values show the balanced accuracy of the shared EEGJev D3 branch on 22 tasks by window length. Dashes mark unavailable tasks.
\textbf{d}, Mean absolute bin error for models covering all 22 tasks.
\textbf{e}, The score averages the rate of refusing when no option fits and the rate of answering when a correct option is present. The full-label check also scores every known class.
\textbf{f}, Matched differences between the shared refusal head and the listed confidence rules. The right column gives each comparison's task count. Whiskers in \textbf{b,f} give paired 95\% intervals from 10,000 resamples of tasks within sources, conditional on the evaluated participants. The scores are from the test split.}
\label{fig:paired-bins-refusal}
\end{figure}

For D2, the joint score gives equal weight to observed answer combinations and requires both questions about the same window to be answered correctly (Figure~\ref{fig:bench}b).
On test, EEGJev, the ordinary classifier and task-specific EEGNet scored 39.3\%, 41.0\% and 44.2\%, respectively (Figure~\ref{fig:paired-bins-refusal}a).
Scoring the questions individually gives balanced accuracies of 61.1\%, 62.2\% and 63.0\% on test and
62.9\%, 64.3\% and 65.3\% on validation, in the same model order.
On the 59 matched tasks, the per-question difference between EEGJev and EEGNet is $-2.0$ percentage points on test
(95\% interval $-3.7$ to $-0.2$, Figure~\ref{fig:paired-bins-refusal}b) and $-2.4$ on validation
($-4.0$ to $-0.9$). The corresponding test difference between the ordinary classifier and EEGNet is $-0.8$ points
($-2.2$ to $+0.6$).

\subsection{Refusal when the correct option is absent}
\label{sec:res-refusal}

EEGJev's D4-trained refusal head (Act) scored 59.2\% on test and 58.6\% on validation
(Figure~\ref{fig:paired-bins-refusal}e). On matched test tasks it scores 4.8 points above the confidence threshold
on its own predictions ($+4.2$ to $+5.4$) and 6.2 above EEGNet's ($+5.1$ to $+7.3$, Figure~\ref{fig:paired-bins-refusal}f).
Confidence rules for NeuroLM, METIS and LEAF score close to 50\%, as does a rule that makes the same decision throughout each group.
On their respective matched tasks, Act scores 7.8 points above NeuroLM's rule ($+6.0$ to $+9.6$) and 9.2 above METIS's
($+7.7$ to $+10.8$). NeuroLM's threshold was calibrated on D4 validation after training on D1--D3.

For tasks seen in training, the full-label check scores all answers of the task and refuses when the best one is
not offered. On the same 31 D4 test tasks, this rule gives 65.4\% for EEGJev and 66.7\% for EEGNet
(Supplementary Note~\suppref{app:knownrules}).

\subsection{Inference rules for known tasks and ordered-bin scores}
\label{sec:res-d2d3d4}

For tasks seen in training, averaging the log-probabilities of EEGJev and the ordinary classifier with equal weights gives
65.2\% on test D1 and 63.2\% on test D2. This is 1.4 points above EEGJev on D1 ($+1.0$ to $+1.8$) and 2.1 points
on D2 ($+1.4$ to $+2.9$), and it is not distinguishable from EEGNet on D1 ($+0.7$, $-0.5$ to $+1.9$) or on D2 ($+0.2$,
$-1.3$ to $+1.6$). Averaging and the full-label check were fixed before v1.3 test evaluation and applied to the saved test outputs without additional fitting (Supplementary Note~\suppref{app:knownrules}). Both apply only to tasks with known training labels.

Figure~\ref{fig:paired-bins-refusal}c,d shows D3 accuracy by task and mean error in bins. The D3 branch of EEGJev reaches 41.5\% on test and EEGNet 42.5\%, a paired difference of
$-1.0$ points ($-2.8$ to $+0.8$). The bin heads of the ordinary classifier reach 42.4\%. On the 14 D3 tasks
covered by LaBraM, the D3 branch scores 34.0\% and LaBraM 39.7\% (Figure~\ref{fig:bench-model-overview}c).
The mean bin distance is
0.80 for the D3 branch, 0.83 for the ordinary bin heads and 0.75 for EEGNet. On validation the branch is at 41.3\% and
EEGNet at 42.7\%. Without the branch, the decision head answers D3 at 35.3\% on test, 7.2 points below EEGNet
(Supplementary Note~\suppref{app:d3}), so the binned questions are matched only with the further-trained branch.

\subsection{Questions over two windows and long windows}
\label{sec:res-t1t2}

\begin{figure}[!tbp]
\centering
\includegraphics[width=\textwidth]{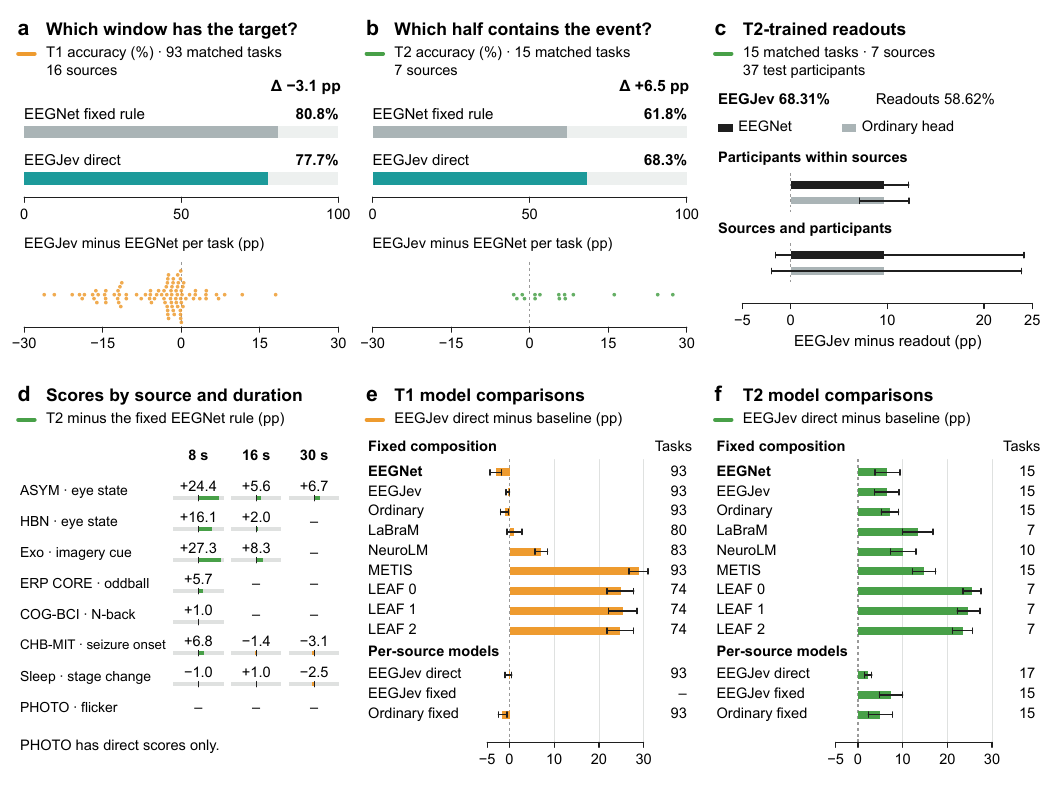}
\par
\caption[Temporal decisions across questions and sources.]{\textbf{Temporal decisions across questions and sources.}
\textbf{a,b}, Bars give source-balanced accuracy on matched T1 and T2 tasks, respectively. Each dot shows the difference between EEGJev and the EEGNet rule on one task. EEGJev trained on temporal labels. The fixed rules compare D1 scores between windows for T1 and combine successive one-second D1 predictions for T2.
\textbf{c}, EEGJev and both readouts trained on the same T2 labels. The readouts use time-ordered one-second class probabilities. Bars show paired source-balanced differences averaged across three seeds on 15 test tasks from seven sources and 37 participants. Whiskers give 95\% intervals from 10,000 paired resamples of participants within fixed sources, or of sources and participants.
\textbf{d}, Signed differences between EEGJev and the fixed EEGNet rule by source and duration. Dashes mark unavailable comparisons. PHOTO has direct scores for two durations.
\textbf{e,f}, Paired differences between shared EEGJev and other models on T1 and T2, respectively. Whiskers give 95\% intervals from 10,000 paired resamples of tasks within sources. Counts give the number of matched tasks in each comparison. The comparisons use fixed compositions except for the per-source direct head, which also trained on temporal labels.}
\label{fig:temporal}
\end{figure}

Figure~\ref{fig:temporal} compares direct T1 and T2 answers with fixed rules that combine D1 outputs
(see ``\nameref{sec:baselines}'').

On T1, EEGJev reaches 77.7\% on test and on validation. Composing the answer from its own D1 outputs gives 78.1\%, so
the native answer is not better than the composition here. The composition from EEGNet reaches 80.8\% and is 3.1 points above EEGJev
(interval 1.8 to 4.4), and the ordinary classifier is 1.1 points above it (0.3 to 2.0). LaBraM (77.5\%) is not
distinguishable from EEGJev on the 80 tasks it covers, and NeuroLM is 7.1 points lower.

With the same early-versus-late training labels, EEGJev read the long window directly and scored 68.31\% on 15 matched T2 test tasks from seven sources (Figure~\ref{fig:temporal}c).
Trained readouts of the frozen one-second EEGNet and shared ordinary classifier both scored 58.62\% using time-ordered class probabilities. Using mean probabilities over the window, they scored 53.21\% and 52.63\%, respectively.
EEGJev led the ordered EEGNet readout by 9.69 points. The 95\% interval was $+7.43$ to $+12.22$ points when resampling the 37 test participants within the seven sources and $-1.55$ to $+24.16$ points when also resampling sources.

We also compared EEGJev with fixed rules that combined consecutive one-second D1 predictions without fitting to T2 labels (Figure~\ref{fig:temporal}b,d,f). EEGJev reaches 69.8\% on
test and 67.1\% on validation. Every composition is lower. The composition from
EEGJev's own D1 outputs reaches 61.8\%, 6.5 points below the native answer ($-9.2$ to $-3.7$), the ordinary classifier
61.2\% ($-7.1$, $-9.1$ to $-5.2$) and EEGNet 61.8\% ($-6.5$, $-9.4$ to $-3.8$). The best composition, from the
per-source ordinary classifier, is 4.9 points lower ($-7.7$ to $-2.3$). On validation every composition is lower as well, with
EEGNet 4.3 points below EEGJev ($-7.5$ to $-1.0$). These comparisons use fixed rules for 15 of the 17 T2 tasks.

\subsection{Shared and per-source encoders}
\label{sec:res-persource}

Training one encoder per source changes little. On test D1, EEGJev with per-source encoders is 0.2 points above the
shared EEGJev ($-0.6$ to $+1.0$) and the per-source ordinary classifier 0.9 points above it (0.0 to $+1.8$). The
per-source EEGJev is higher on D4 (60.5\%) and lower on T2 (67.5\%), and it loses more under the held-out phrasings.
Supplementary Note~\suppref{app:persource} gives the other comparisons.

\subsection{Held-out tasks}
\label{sec:res-heldout}

A question-reading model can in principle answer a question it was not trained on, which a classifier without an
output layer for it cannot. We removed one task at a time from all training, including step 1 of the encoder, its D1
rows, the D2 rows paired with it, its D3 and D4 rows, its T1 pairs and its classifier layer. We then asked the
retrained model the task's validation questions, which no part of the new model had used. Eye state of vrnf is a
concept that other sources also ask about. Face gender of asym appears in no other source (several sources ask the
gender of the participant, which is a different attribute).

Figure~\ref{fig:controls-transfer}c,d shows the two held-out tasks. Supplementary Table~\suppref{tab:heldout} gives the individual training-seed scores. With vrnf eye state removed, EEGJev answers it at 78.9\% balanced
accuracy (seeds 76.7 to 80.8\%, AUROC 0.86), against 48.6\% with shuffled EEG and 83.1\% for the model trained on
the task. The per-source EEGJev reaches 81.2\% (75.8 to 85.0\%) against 87.2\% in training. Eye state learned from
other sources thus carries over to a source whose eye-state questions were never seen. Face gender does not transfer.
EEGJev reaches 51.7\% (46.5 to 60.3\%, AUROC 0.53) and the per-source EEGJev 47.9\%, at the level of shuffled EEG
(49.7 and 50.1\%), while the models trained on the task reach 82.1 and 82.5\%. One seed of EEGJev is at 60.3\% and
the other two are below 50\%. The evaluation is small (240 questions from five validation persons for vrnf, 3,909 from
14 persons for asym), and the ordinary classifier, EEGNet and LaBraM are not compared here because they have no output
for a task removed from training.

\subsection{Datasets no model has seen}
\label{sec:res-unseen}

On left-or-right-hand motor imagery in two datasets that no model had been trained or evaluated on,
Weibo2014~\cite{yi2014weibo} with 10 persons and Dreyer2023A~\cite{dreyer2023large} with 60, EEGJev stays at chance
(49.6\% to 52.4\%). The ordinary classifier (53.1\% to 58.1\%) is above it in all six conditions, by 2.2 to 7.0
points with intervals that exclude zero. Supplementary Note~\suppref{app:unseen} describes the setup and few-shot calibration.

\subsection{Response time for individual questions}
\label{sec:res-response-time}

EEGJev and NeuroLM-B answered the same 192 binary D1 validation questions (Exo walking transitions and LEMON eye state) one at a time, with checkpoints from three training seeds and 576 timed calls per model per platform.
On one CPU core, the 95th-percentile response times were 394.97\,ms for EEGJev and 3,677.34\,ms for NeuroLM-B. On one GPU they were 9.60 and 24.99\,ms, respectively. On each platform, EEGJev returned an offered option in all 576 calls and NeuroLM-B in 554.
Timing started with preprocessed EEG in host memory, included text processing and answer return, and counted unparseable responses as infinite time to an offered option (Supplementary Note~\suppref{app:compute}).

\section{Discussion}
\label{sec:discussion}

EEGJev-Bench measures two decisions separately. Does the answer match the requested attribute, and is its value correct?
On attribute pairs seen in training, EEGJev selected the requested attribute with 98.0\% to 100\% accuracy across 13 test sources.
Balanced accuracy for the specific answer ranged from 40.2\% to 85.0\% (Figure~\ref{fig:controls-transfer}a).
Task-specific EEGNet achieved higher joint accuracy on paired questions, and EEGJev's refusal head outperformed EEGNet's confidence rule.
EEGJev scored the offered answers in one pass and had a shorter 95th-percentile response time than NeuroLM-B on the same 192 validation questions answered individually, on both a single CPU core and a GPU.

Golan et al.~\cite{golan2023languagejudgements} used paired sentences to reveal failures that aggregate language-model scores could miss.
Gu et al.~\cite{gu2026cardiacsensing} evaluated candidate selection in cardiac question answering, and EEGJev-Bench holds the EEG and answer choices fixed to test whether a decoder follows a change in the requested attribute.

Most sources taken from UniEQA have one to three test participants. Bootstrap intervals over tasks condition on
these participants. Validation is used for model selection or calibration, and the D1 gap between EEGJev and EEGNet
is larger on validation than on test.
The long-window result rests on 17 tasks from 8 sources, and the event times of T2 are instruction, cue or scoring
times. The refusal head scores 59.2\%, 9.2 percentage points above the 50\% baseline for a rule that ignores the EEG. Letter codes
in place of the candidate names bring EEGJev down to the level of the controls without EEG, so it depends on the
wording of the candidates. The held-out task results cover one concept seen in other sources and one attribute that
no other source asks, and they show that transfer is possible, not how often it occurs. The motor-imagery transfer fills
missing electrodes by interpolation, and the training data contain only two persons who imagine left and right hand
movements. The decision-head stage of EEGJev trains 806.8 million parameters. Baselines use the benchmark's splits and preprocessing.
The fixed T1 and T2 rules combine D1 outputs. The supervised T2 readouts use the same T2 labels as EEGJev
(Figure~\ref{fig:temporal}). METIS and LEAF use released weights. Comparisons with them involve differences in model design and training conditions.

EEGJev lets users specify what to decode from an EEG recording through a question and returns a probability for each offered answer in one pass.
EEGJev-Bench holds the recording fixed while varying the question or the availability of the correct option, so researchers can assess question following, answer accuracy and refusal separately.

\section{Methods}
\subsection{EEGJev-Bench}
\label{sec:benchmark}

EEGJev-Bench asks multiple-choice questions about attributes such as the stimulus shown, the sleep stage or the participant's diagnosis.
All results use data version v1.3.

\subsubsection{Sources and windows}
\label{sec:bench-sources}

We use recordings from the original releases of all 19 sources (Figure~\ref{fig:bench}, Supplementary Note~\suppref{app:bench-sources},
Supplementary Table~\suppref{tab:bench-sources}). For 12 sources from Unified\_EQA, the public EEG question-answering
release in UniEQA~\cite{unieqa2026}, we locate each of its segments in the original recording. The seven added sources are
COG-BCI~\cite{hinss2023cogbci}, HBN-EEG~\cite{alexander2017hbn,shirazi2024hbneeg}, a verbal working memory
dataset~\cite{pavlov2020vwm}, the Dortmund Vital Study~\cite{getzmann2024dortmund}, ERP CORE~\cite{kappenman2021erpcore},
MPI-LEMON~\cite{babayan2019lemon} and a resting-state depression dataset~\cite{cavanagh2021deprestds}. Seventeen sources have scored tasks. The bciciv and is sources have three training persons each and no
test persons in this version, so they serve as training material only.

Short windows last 1, 2 or 4\,s. An event label gives one window from the event, and a state or person label gives at
most eight non-overlapping windows per segment. Long windows of 8, 16 or 30\,s are cut around an event for the T2
questions.

All recordings come from public datasets, and no new data were recorded for this study. For 15 of the 19 sources,
the original publications or dataset pages state an ethics approval and the consent of the participants. Those of
BCI Competition IV, CHB-MIT and Sleep-EDF state neither, and those of the depression dataset state consent without
naming a review board.

\subsubsection{Question types}
\label{sec:bench-questions}

D1 asks one attribute of one window. Its options are all train values of that attribute for the same source and
window length, so they depend on the cell (source, attribute, window length) and not on the window. Each attribute has
three phrasings, and a D1 row uses one of them. Each attribute also has two phrasings that occur in no training text,
and every D1 test row is repeated with both of them (D1h).

D2 asks two attributes of the same window, each in all three phrasings. The pair is chosen so that neither attribute
determines the other over the train segments of the source. D2 is scored per attribute, and a joint score counts a
window as correct only when both attributes are answered correctly.

D3 asks the model to choose among ordered categories. The questions cover MMSE by clinical cut points
($\geq 24$, 18--23, $<18$), age in the Dortmund, HBN and LEMON sources, sleep stage by level (Wake, N1, N2, N3), n-back
load, the level of the MATB task and the foreground contrast in HBN. During benchmark construction, we merged two
LEMON age bins (55--65 and 65--70 years) after the test label counts showed that the younger bin was empty
(Supplementary Note~\suppref{app:bench-quality}). D3 is scored by balanced accuracy and by the mean absolute distance
between the chosen and the true bin.

D4 tests refusal. For each D1 question with at least three options, D4 adds a refusal row, in which the correct option
is removed, and a control row, in which one wrong option is removed and the correct answer remains. Both rows have one
option fewer than the D1 question, so the number of options does not reveal the correct decision.

T1 shows two windows of the same person and of the same length and asks which of them carries a given label of a D1
task. Each pair gives four linked views, with the two windows in either order and each of the two labels asked. A
window enters at most one pair of a task. T2 asks whether an event lies in the first or the second half of a long
window. The event is placed near one quarter or three quarters of the window, with a jitter of one sixteenth of its
length, and the two placements of an event form one linked group. The events are changes of eye state, seizure onsets,
sleep stage changes, cued transitions, changes of flicker frequency and targets in a stimulus stream. Their labels
are instruction, cue or scoring times, not measured neural onsets.

The train and validation persons have 940,604 short windows and 6,728,339 D1 to D4 questions. The test persons have
117,571 short windows and 878,827 questions, and D1h adds 488,786 rows. The scored T1 tasks have 557,933 train pairs,
94,924 validation pairs and 80,748 test pairs, and the scored T2 tasks have 64,465, 12,315 and 3,368 event groups.

\subsubsection{Splits}
\label{sec:bench-splits}

In the 17 scored sources, participants belong to only one split. For the seven added sources, 15\% of the persons form the test split and 15\% of the rest
the validation split, drawn with fixed seeds. For the sources from UniEQA, test and validation persons were drawn with
fixed seeds in the numbers of the earlier data version, so most of these sources have few test persons (ten for
dementia and photo, three for asym and sleep, two for chbmit, mom, surprise and vrnf, one for exo and gal). Photo and
dementia record the same participants and are split jointly. The added sources have between 4 test persons (cogbci)
and 91 (dortmund). Results are therefore reported per task and with the number of test persons in mind.

\subsubsection{Data quality rules}
\label{sec:bench-quality}

In a study of six EEG datasets, changing preprocessing alone altered up to 42\% of trial-level
predictions~\cite{hou2026samebrain}. We use one signal pipeline to construct the benchmark windows.
We read each recording from its original file using a fixed channel list. For ERP CORE, we first subtract the
mean across the 30 scalp channels at each time point. We apply a notch filter at the mains frequency where a
peak is present, resample the whole recording to 200\,Hz and standardize each channel within each window.
We exclude windows that start in the first 2\,s, end in the last 0.5\,s, lie within 0.1\,s of a discontinuity marker
or contain flat or dead channels. We correct labels against the original data, for example by restoring the official
bciciv motor imagery labels and cutting each gal window from its own lift (Supplementary Note~\suppref{app:bench-quality}).

\subsubsection{Evaluation protocol}
\label{sec:bench-eval}

Each row is scored with its own options in the given order. A task is a cell of one question type. The main metric
of D1, D1h, D2 and D3 is balanced accuracy over answer labels, computed within each task. A task is scored only if
every offered answer has at least 10 windows on validation and on test, and this task list was fixed on 15 November
2026 from the label counts alone, before any model was trained on this data version. It has 109 D1 tasks in 17
sources, 59 D2 tasks in 10 sources, 22 D3 tasks in 6 sources and 31 D4 tasks in 11 sources. Scores are averaged with
equal weight for every source and equal weight for the tasks within a source (the source-balanced mean), so that a
source with many tasks does not dominate. D3 also reports the mean absolute bin distance. The D2 joint score averages
the nine combinations of phrasings of a window and is balanced over the combinations of the two answers.

In the fixed-option experiment, we count a question pair as correct only when both answers are correct and average the scores within each pairing condition, across conditions within each source and then across sources.

D4 is scored by a group-balanced criterion. A group holds the D4 rows that share source, attribute, window length and
the unordered set of offered labels. Within a group, the criterion is the mean of the refusal recall on refusal rows
and the answer rate on control rows, and a task's score is the mean over its groups that contain both kinds of rows.
A rule that does not look at the EEG cannot exploit which label is missing, since every decision that is constant
within a group scores exactly 0.5. Every model can refuse by thresholding its calibrated maximum option probability.
Per model, one temperature is fitted on validation D1 by negative log-likelihood and one threshold, shared across
sources, is chosen on validation D4. Models with their own refusal output are also scored with it. Calibration is
reported as the expected calibration error over 15 bins (ECE) of the D1 answers after temperature scaling.

T1 and T2 are scored by accuracy. The linked views of a pair or an event group are averaged first, then persons,
tasks and sources, so that the four views of a pair do not count as four samples. A T1 or T2 task is scored when
validation and test each have at least ten pairs or event groups, which leaves 93 T1 tasks in 16 sources and 17 T2
tasks in 8 sources. Chance is 50\% for both.

\subsubsection{Controls without EEG}
\label{sec:bench-controls}

Three controls receive no information from the EEG of the window being asked about (an answer prior, a text-only
\href{https://huggingface.co/convaiinnovations/laya/tree/aa8c91ca088ec597df95a0d1c76b3063cb2ae5e8}{Laya checkpoint}~\cite{laya_code} and shuffled EEG). On D1 they reach 43.5\% to 43.8\% on validation and
test, the level of the answer prior (Supplementary Note~\suppref{app:bench-controls}, Supplementary Table~\suppref{tab:bench-nosignal}).

\subsection{EEGJev}
\label{sec:model}

EEGJev answers a multiple-choice question about EEG. It receives one window, or two windows for T1, the question text
and the offered candidates. It returns one score per candidate and, from a separate output, whether it should refuse.
The candidate set can change at run time, and nothing is generated or parsed (Figure~\ref{fig:model}).

\paragraph{EEG encoder.}
The encoder is CBraMod~\cite{wang2025cbramod} with the official pretrained weights. It reads each source's native
channels at 200\,Hz in patches of one second, so a window of $C$ channels and $L$ seconds gives $C \times L$ tokens of
width 200. In a first step, one encoder is fine-tuned on the D1 labels of all 19 sources together. Each source and
window length has its own hidden layers, namely the two-layer head that CBraMod places on the flattened tokens
($200CL$ to 800 to 200 units, with ELU and dropout 0.1), and each cell has its own linear output layer. The encoder
is then frozen. The same step-1 model with its per-cell output layers is the ordinary classifier described in
``\nameref{sec:baselines}''.

\paragraph{Decision head.}
The decision head is a small bidirectional transformer that reads one sequence per question. The sequence holds the
window's $C \times L$ CBraMod tokens, one spatial token, and the text tokens of the question and of each candidate. Each
CBraMod token is projected to width 256 and receives a channel and a patch embedding. The spatial token summarizes the
whole window. New hidden layers of the same shape as the step-1 heads, one set per source and window length, map the
flattened CBraMod tokens to 200 units, and a linear layer and a learned embedding turn the result into one more token.
CBraMod has no vocabulary, so the text is read by a frozen MPNet encoder (\href{https://huggingface.co/sentence-transformers/all-mpnet-base-v2}{all-mpnet-base-v2}~\cite{song2020mpnet}), whose 768-dimensional token outputs are projected to width 256. Segment and question-type embeddings
mark which part of the sequence a position belongs to, and each candidate has a marker position. The head has two
pre-norm layers with four attention heads, a feed-forward width of 1,024 and dropout 0.1. Every position attends to
every other position, so each candidate is scored in the context of the others. A two-layer scorer reads each marker
and returns the candidate's logit, and a softmax over the offered candidates gives the answer probabilities. This
joint scoring of candidates at markers follows the typed decision model Laya~\cite{laya_code}. EEGJev uses none of
Laya's weights.

\paragraph{Two windows and long windows.}
The same head answers T1 and T2. For T1, the tokens of both windows enter the sequence, and a learned embedding marks
each token as belonging to window A or window B. The two candidates name the windows, and the tokens of a candidate
attend to the EEG tokens of its own window only. For T2, the long window is cut into consecutive blocks of at most
4\,s (1\,s for erpcore, which has hidden layers for 1\,s only), the frozen encoder reads each block separately with the hidden layers of its source and length, and a learned
embedding gives the position of each block. The tokens of the candidate for the first half attend to the EEG tokens
of the blocks in the first half only, and likewise for the second half. A block that spans the midpoint belongs to
both halves. No new encoder layers are added for long windows.

The decision-head stage trains 806.8 million parameters. Its 52 sets of spatial-token hidden layers contain
804.7 million (99.74\%), and its two transformer layers together contain 1.58 million.
With the frozen CBraMod and MPNet, EEGJev is therefore not a small model (Supplementary Note~\suppref{app:compute}).

\paragraph{Refusal output.}
A separate Act head decides whether to answer. It is a two-layer network (260 to 256 to 2 units) that reads the first
output position of the decision head and four features of the answer distribution, namely the top probability, the
gap between the top two probabilities, the entropy divided by $\log K$ and the number of offered candidates $K$
divided by 255. EEGJev refuses when the Act head's refusal output is the larger of its two outputs.

\paragraph{Training.}
With the encoder frozen, one decision head is trained on all question types except D3 and D4. The 3,258,618 D1 and
D2 questions of the training persons are used for 18,900 updates of 64 questions with soft cross-entropy over the
offered candidates. Options are shuffled, 25\% of the questions offer a random subset of their options, and 10\%
replace the option texts by letters whose mapping is stated in the question. For a D1 question whose source asks
another attribute at the same window length, a quarter of the draws offer the labels of both attributes together, so
that the question decides which attribute is asked. Mixed into the same schedule are 10,463 updates on T1 pairs and
1,913 updates on T2 event groups, 31,276 updates in all. The Act head is trained afterwards, with the decision head
frozen, on pairs of D4 training rows, to refuse on refusal rows and to answer on control rows. The decision head
sees no D3 questions, and its own D3 balanced accuracy is 34.1\% on validation. A D3 branch therefore continues from
the trained head on D3 training questions mixed with D1 and D2 questions and answers the D3 rows only. All D1, D2, D4,
T1 and T2 results use the main head. Three seeds are trained, and each seed has its own step-1 encoder.
Supplementary Note~\suppref{app:training} summarizes the training schedule.

\subsection{Experimental setup}
\label{sec:experiments}

\subsubsection{Baselines}
\label{sec:baselines}

Baselines use the benchmark's windows, splits and scorer on the tasks they cover. Paired comparisons use rows that
both models score. These scores describe performance under our benchmark protocol. EEGNet~\cite{eegnet2018}
and LaBraM~\cite{labram2024} are trained for
each D1 and D3 task. LaBraM has no models for chbmit, hbn and sleep, whose channels its channel table does not cover.
The ordinary classifier is EEGJev's step-1 model (see ``\nameref{sec:model}''). It does not read the question. These
three answer a D2 row with the model or output layer for the requested attribute. The default D4 refusal rule uses
a confidence threshold. We report the full-label check separately where available.

Three baselines read the question. NeuroLM~\cite{neurolm2024} is instruction-tuned on the D1 and D2 training
questions of the benchmark without text-only batches, followed by a short stage on D3 questions, and has no input for
hbn and sleep. METIS~\cite{metis2026} and LEAF~\cite{leaf2025} are run from their released checkpoints without
training on the benchmark. METIS scores the option letters in one forward pass. LEAF covers 12 of the 17 scored
sources and is reported at its three
instruction levels. As comparison rows, we also build the ordinary classifier and EEGJev on per-source encoders.
Supplementary Note~\suppref{app:baselines} describes the training and inference settings.

The fixed comparisons in Figure~\ref{fig:temporal} combine each baseline's D1 outputs without fitting to T1 or T2 labels.
For T1, the composed answer is the window to which the model's D1 output gives the higher probability of the asked
label. For T2, the model's one-second D1 output is computed on every consecutive second of the long window. For a
change of state, the answer is the half that holds the second at which the mean output before and after differs most,
and for a target in a stream it is the half that holds the second with the highest target probability. The photo
events have no matching D1 attribute, so 15 of the 17 T2 tasks can be composed. These rules were fixed before any
model output existed. The same compositions applied to EEGJev's own D1 answers serve as a control.

For the supervised T2 comparison, we kept the one-second EEGNet and shared ordinary classifiers frozen and fitted a logistic regression for each task and seed. The input was either the class probabilities concatenated in time order or their mean across the window. We standardized each feature using the training split and selected the inverse regularization strength ($C=0.01, 0.1, 1$ or $10$) on validation, choosing the smaller value in a tie. The selected readouts were then applied to the test split.

\subsubsection{Protocol}
\label{sec:protocol}

The tables average three seeds for each trained model. We use validation for checkpoint selection where applicable
and to fit temperatures and D4 thresholds. NeuroLM uses the checkpoint after its final 300-step stage (Supplementary Note~\suppref{app:baselines}).
The model, training recipe, task list and primary metrics were frozen before the v1.3 test.
Further v1.3 analyses include fixed-option question switching, participant and source resampling, supervised T2 readouts and full-label refusal rescoring.
Supplementary Notes~\suppref{app:d3}, \suppref{app:knownrules} and~\suppref{app:figure-scores} describe D3 head scores, averaging and the paired D4 comparisons.
Most of the earlier sources have few test participants
(see ``\nameref{sec:bench-splits}'').

Differences between models are computed per task on the rows that every model in the comparison scored. They are
given as the source-balanced mean of the paired task differences with a 95\% interval from a bootstrap over tasks
within sources (10,000 resamples). These task-based intervals condition on the participants evaluated. For the D1 comparison of EEGJev and EEGNet,
we also computed a paired 95\% percentile interval from 10,000 resamples of participants within sources, keeping
each participant's windows together across both models and all three seeds. The same draw was used for participants shared by dementia and photo. For matched T2 readouts, we also used 10,000 paired resamples to estimate intervals. One analysis resampled participants with matched T2 data within fixed sources. The other resampled sources first and then drew participants from each selected source's test pool. Both placements of an event stayed with their participant. The scores in the tables are each model's own source-balanced means over the tasks
it covers, so they can differ slightly from the paired differences.

Supplementary Note~\suppref{app:compute} reports model sizes and response times.

\backmatter
\section*{Data availability}
Supplementary Table~\suppref{tab:bench-sources} lists the source datasets and providers. Benchmark questions, labels and the window index remain private.
\par\vspace{-4pt}
\section*{Code availability}
A private reviewer package provides code, weights and a synthetic CPU example. The
\href{https://github.com/dengzhe-hou/EEGJev/tree/main/paper/v1.3}{project repository}
contains detailed training settings, data quality checks, per-seed results and figure source data.

\FloatBarrier
\setlength{\bibsep}{6pt plus 1pt minus 1pt}
\renewcommand{\bibsection}{%
  \section*{\refname}\addcontentsline{toc}{section}{References}}
\bibliography{references,refs_draft_benchmark_20261026,refs_draft_related_20261026,refs_paper_20261113}

\raggedbottom
\section*{Supplementary Information}
\addcontentsline{toc}{section}{Supplementary Information}
\begingroup
\linespread{0.95}\selectfont
\newcounter{suppnote}
\makeatletter
\newcommand{\EEGJevSupplementaryHeading}{%
  \@startsection{section}{1}{\z@}{-8pt plus -2pt minus -1pt}{5pt}{\sectionfont}}
\makeatother
\newcommand{\suppnote}[1]{%
  \refstepcounter{suppnote}%
  \EEGJevSupplementaryHeading*{Supplementary Note \thesuppnote. #1}%
  \addcontentsline{toc}{subsection}{Note \thesuppnote. #1}}
\setcounter{table}{0}
\renewcommand{\tablename}{Supplementary Table}
\renewcommand{\theHtable}{supplementary.\arabic{table}}

\suppnote{Sources and windows}
\label{app:bench-sources}

Table~\ref{tab:bench-sources} lists the 19 sources of data version v1.3. Twelve of them are sources of the
Unified\_EQA part of UniEQA, whose published question inventory lists 13 sources. We excluded tdcs because its release~\cite{mensah2025tdcsds} describes the hand-muscle recordings as EMG,
although the channel tables mark them as EEG. UniEQA does not publish
its segmentation code, so we inferred what each segment is from segment durations, segment order and the original
event files, and we located every segment in its original recording. The exo files come from OpenNeuro ds006940,
which OpenNeuro has since removed as a duplicate of ds007788. We cite ds007788. The seven added sources are read
from their original releases with segments that we defined from the documented event codes.

\begingroup
\EEGJevCompactTableFont
\setlength{\tabcolsep}{1.4pt}
\setlength{\LTcapwidth}{\linewidth}
\setlength{\LTpre}{6pt}
\setlength{\LTpost}{6pt}
\begin{longtable}{@{}lJ{88pt}J{106pt}rrr@{}}
\caption{Sources of EEGJev-Bench (data version v1.3). Accession numbers refer to OpenNeuro unless another archive is
named. Ch.\ is the number of EEG channels used, in the native montage of each source. For chbmit these are 18 of the
bipolar derivations in the files. Two CHB-MIT recording cases belong to one person and both are in training.
Persons and windows are given as train / validation / test, and windows count all
lengths of 1, 2 and 4\,s. Dementia and photo record the same persons. The bciciv and is sources have no test persons,
and their validation part is split by segment. The seven added sources (cogbci, deprest, dortmund, erpcore, hbn,
lemon, vwm) are named by their archive identifiers. At the request of the vrnf dataset authors, we also cite
MNE-BIDS~\cite{appelhoff2019mnebids}.}
\label{tab:bench-sources}\\
\toprule
\EEGJevTableHead \textbf{Key} & \textbf{Original dataset} & \textbf{Attributes asked} & \textbf{Ch.} & \textbf{Persons} & \textbf{Windows} \\
\midrule
\endfirsthead
\multicolumn{6}{@{}l}{\textbf{Supplementary Table \thetable\ (continued)}}\\
\toprule
\EEGJevTableHead \textbf{Key} & \textbf{Original dataset} & \textbf{Attributes asked} & \textbf{Ch.} & \textbf{Persons} & \textbf{Windows} \\
\midrule
\endhead
\midrule
\endfoot
\endlastfoot
asym & EEG-Asymmetries, ds007172~\cite{reinke2026asymmetries,reinke2026asymds} & task, emotion, face gender, face race, word category, response hand, eyes & 32 & 82 / 14 / 3 & 68,065 / 11,623 / 2,492 \\
bciciv & BCI Competition IV, set 1~\cite{blankertz2007bbci,bciiv2012} & training material only (imagined movement) & 59 & 3 / seg. / 0 & 3,570 / 630 / 0 \\
chbmit & CHB-MIT~\cite{shoeb2009thesis,guttag2010chbmit,pollard2026physionet} & seizure or pre-seizure & 18 & 18 / 3 / 2 & 6,929 / 623 / 1,284 \\
cogbci & COG-BCI, Zenodo 7413650~\cite{hinss2023cogbci,hinss2022cogbcids} & task, eyes, n-back load, n-back target, MATB level, flanker congruency, response outcome & 62 & 21 / 4 / 4 & 96,222 / 18,327 / 18,313 \\
dementia & AD/FTD rest EEG, ds004504~\cite{miltiadous2023dataset,miltiadous2023dicenet,miltiadous2024dementiads} & diagnosis, gender, and MMSE in D3 & 19 & 66 / 12 / 10 & 39,049 / 6,785 / 5,546 \\
deprest & EEG: Depression rest, ds003478~\cite{cavanagh2019depression,cavanagh2021deprestds} & eyes, gender & 62 & 88 / 16 / 18 & 16,775 / 3,074 / 3,460 \\
dortmund & Dortmund Vital Study, ds005385~\cite{gajewski2022dortmund,getzmann2024dortmund,wascher2025dortmundds} & eyes, gender, session part, and age in D3 & 64 & 439 / 78 / 91 & 56,340 / 10,176 / 11,808 \\
erpcore & ERP CORE, OSF 9f5w7~\cite{kappenman2021erpcore,luck2020erpcoreds} & task, stimulus category, scrambling, oddball target, arrow direction, congruency, search side, word relation, response outcome, gender & 30 & 29 / 5 / 6 & 39,440 / 6,800 / 8,159 \\
exo & Exoskeleton, ds007788~\cite{sarkar2026exo,sarkar2026exods} & exoskeleton state, transition & 60 & 5 / 1 / 1 & 60,891 / 9,082 / 11,942 \\
gal & WAY-EEG-GAL~\cite{luciw2014wayeeggal,luciw2020wayeeggalds} & weight, friction & 32 & 9 / 2 / 1 & 6,364 / 1,444 / 712 \\
hbn & HBN-EEG, Release 1~\cite{alexander2017hbn,shirazi2024hbneeg,shirazi2025hbnr1ds} & task, eyes, foreground contrast, and age in D3 & 120 & 99 / 17 / 20 & 53,123 / 7,452 / 10,328 \\
is & Inner speech, ds006033~\cite{simistira2025innerspeech,simistira2025innerspeechds} & training material only (word, task or rest) & 63 & 3 / seg. / 0 & 3,157 / 390 / 0 \\
lemon & MPI-LEMON, preprocessed EEG~\cite{babayan2019lemon,babayan2018lemonds} & eyes, and age in D3 & 17 & 147 / 26 / 30 & 59,838 / 10,519 / 12,197 \\
mom & Object monitoring, ds006768~\cite{lowe2026lapses,lowe2025momds} & trial type, task relevance & 64 & 24 / 4 / 2 & 67,962 / 11,197 / 5,599 \\
photo & AD/FTD photic, ds006036~\cite{ntetska2025photo,ntetska2026photods} & stimulation or rest, eyes & 19 & 66 / 12 / 10 & 7,142 / 1,186 / 1,031 \\
sleep & Sleep-EDF Expanded~\cite{kemp2000sleep,sleepedfx2018,pollard2026physionet} & sleep stage & 2 & 82 / 15 / 3 & 101,406 / 16,566 / 3,840 \\
surprise & Surprise, ds006394~\cite{leong2026surprise,leong2025surpriseds} & trial type, task variant & 16 & 26 / 5 / 2 & 13,230 / 2,067 / 888 \\
vrnf & VR neurofeedback, ds006839~\cite{aguilargonzales2026alpha,aguilargonzales2025vrnfds} & feedback valence, task or rest, eyes & 29 & 29 / 5 / 2 & 9,951 / 1,711 / 687 \\
vwm & Verbal working memory, ds003655~\cite{pavlov2020vwm,pavlov2021vwmcodes,pavlov2023vwmds} & trial phase, gender & 19 & 113 / 20 / 23 & 94,734 / 16,764 / 19,285 \\
\midrule
\textbf{Total} & & & & & 804,188 / 136,416 / 117,571 \\
\bottomrule
\end{longtable}
\endgroup

Windows follow what a label refers to in time. Event labels describe a stimulus, a cue or a trial condition that
starts with the segment. Such a segment gives one window from its start. State and person labels, such as eye state,
sleep stage, seizure or diagnosis, give non-overlapping windows, at most eight per segment, so that long recordings do
not dominate. Windows last 1, 2 or 4\,s (1\,s only for erpcore, whose trials are shorter than 2\,s, and for surprise, and 1 and
2\,s for mom), stay within
their segment and are never padded with zeros. In chbmit, pre-seizure windows are cut backward from the seizure
onset, so all of them lie within 32\,s of it.

The OpenNeuro metadata archived for our source checks identify mom v1.1.0, photo v1.0.6 and dementia v1.0.9.
The dementia recordings and participant table are identical in v1.0.9 and the cited v1.0.8.
The UniEQA release does not specify which OpenNeuro snapshots it used for these three sources.

\suppnote{Data quality rules}
\label{app:bench-quality}

\paragraph{Signal pipeline.}
We use a fixed channel list and order for each source and exclude recordings with missing channels.
For 50, 60 or 100\,Hz, we apply a zero-phase notch filter to the whole recording if the median power ratio
relative to frequencies 3--8\,Hz away is at least 3 in sampled training recordings. This is the only filter.
We drop windows with all channels flat for at least 0.1\,s or any channel flat throughout.
Only ERP CORE is re-referenced, by subtracting the mean of its 30 scalp channels~\cite{kappenman2021erpcore}.
We retain 120 of HBN's channels E1--E128 after dropping eight that are often constant, 62 channels
excluding Cz for COG-BCI, the 17 common channels for LEMON and 62 channels for the depression rest dataset.

\paragraph{Source corrections.}
\begin{itemize}
\item In bciciv, we exclude three artificial recordings~\cite{bciiv2012}. Matching windows to the official
  100\,Hz calibration cues revealed reversed UniEQA labels. Lateralized band power supported the official
  mapping, which we restored.
\item We label the two vrnf rest periods in recording order (open eyes, then closed eyes). Posterior 8--13\,Hz
  relative power was higher in the second period for 33 of 34 checked participants. We omit eye-state questions
  for the participant whose signal showed the reverse pattern.
\item In photo, the original flash markers distinguish stimulation from rest after the flashes stop.
\item In gal, we align each window to its own lift in the original WS files. The recorded surface codes and
  grip-to-load ratios support exchanging the labels silk and sandpaper and retaining suede.
\item In exo, we read the 320 affected segments at 100\,Hz to correct early starts caused by truncation to 99\,Hz.
\item For one chbmit participant, we derive 18 bipolar channels by taking electrode differences in 26
  common-reference segments. Their amplitudes were similar to those of the participant's native bipolar segments.
\end{itemize}

\paragraph{Question eligibility and participant attributes.}
We omit training cells with a single answer class because window length would reveal the answer. Scored tasks
have at least 10 windows per answer in each of validation and test. We merged the LEMON age bins of 55--65
and 65--70 years after checking test label counts and finding no test participant in the younger bin.
We hold age and gender constant across each participant's windows.

\suppnote{Controls without EEG}
\label{app:bench-controls}

The answer prior predicts the most frequent training answer in each cell, giving chance balanced accuracy.
The pretrained \href{https://huggingface.co/convaiinnovations/laya/tree/aa8c91ca088ec597df95a0d1c76b3063cb2ae5e8}{Laya checkpoint}~\cite{laya_code} reads the question and options zero-shot.
The shuffled-EEG control supplies the ordinary classifier with another window from the same source and of the same
duration. Table~\ref{tab:bench-nosignal} compares these controls with the ordinary classifier on matching EEG.

\begin{table*}[htbp]
\centering
\caption{Controls without EEG and the ordinary classifier with real EEG, source-balanced means in \% over the scored
tasks of data version v1.3. The two ordinary-classifier rows are means of three seeds. The answer prior and
text-only Laya do not depend on a seed.
Bold marks the highest score in each column.}
\label{tab:bench-nosignal}
\EEGJevTableFont
\begin{tabular}{@{}lrrrrrr@{}}
\toprule
 & \multicolumn{2}{c}{\cellcolor{EEGJevTableBlue}D1} & \multicolumn{2}{c}{\cellcolor{EEGJevTablePurple}D2} & \multicolumn{2}{c}{\cellcolor{EEGJevTablePurple}D2 joint} \\
\cmidrule(lr){2-3}\cmidrule(lr){4-5}\cmidrule(l){6-7}
\EEGJevTableHead \textbf{Scorer} & \textbf{Val} & \textbf{Test} & \textbf{Val} & \textbf{Test} & \textbf{Val} & \textbf{Test} \\
\midrule
Ordinary classifier, real EEG & \textbf{64.9} & \textbf{64.2} & \textbf{64.3} & \textbf{62.2} & \textbf{44.4} & \textbf{41.0} \\
Ordinary classifier, shuffled EEG & 43.7 & 43.8 & 44.5 & 44.5 & 20.8 & 21.1 \\
Answer prior & 43.5 & 43.5 & 44.4 & 44.4 & 20.8 & 21.4 \\
Text-only Laya, zero-shot & 43.7 & 43.6 & 44.4 & 44.4 & 19.1 & 19.6 \\
\bottomrule
\end{tabular}
\end{table*}

\suppnote{Training details of EEGJev}
\label{app:training}

\noindent\emph{Step 1.}\enspace We fine-tuned for 50 epochs and kept the checkpoint with the highest mean validation balanced accuracy over cells.

\noindent\emph{Decision head.}\enspace Over 50 epochs, we sampled sources in proportion to the square root of their question counts and window lengths in proportion to their counts. We weighted questions by inverse answer frequency within each cell and kept both questions from each D2 window together. D1 draws offering two attributes used up to 12 options and omitted the subset and letter augmentations. We used 557,933 T1 training pairs and 64,465 T2 training event groups. Each sampled pair or group contributed one linked view under the same sampling weights. We selected the checkpoint on D1 and D2 validation questions, with up to 300 per task.

\noindent\emph{Act and D3 heads.}\enspace The frozen decision head supplied features to the Act head, which used up to 2,000 D4 pairs per cell for 11 epochs of 788 updates with 64 rows each. Pairs from 15\% of training windows were held out for epoch selection. The D3 branch continued from the trained head for six epochs of 400 updates, with each batch split evenly between D3 questions and a pool of D1 and D2 questions.

\suppnote{Baselines}
\label{app:baselines}

\paragraph{EEGNet and LaBraM.}
EEGNet~\cite{eegnet2018} uses the official architecture (revision 4a512e5) in PyTorch. Across three seeds, 134 tasks including three unscored bciciv tasks yielded 402 models, each trained for 500 epochs and selected by validation cross-entropy. LaBraM~\cite{labram2024} uses the official 50-epoch TUAB recipe and an effective batch of 512, or the full training set for four smaller tasks. Selection used validation accuracy. The 113 covered tasks yielded 339 models across three seeds. Validation and test inference used deterministic GPU settings.

\paragraph{Ordinary classifier.}
Its D3 bin heads use frozen step-1 features and the same D3 training questions as EEGJev's D3 branch. The per-source variants fine-tune one encoder per source on its own D1 labels and follow the same recipe thereafter.

\paragraph{NeuroLM, METIS and LEAF.}
NeuroLM~\cite{neurolm2024} used five epochs on D1 and D2 without text-only batches, then 300 updates with half D3 questions. All tracks use the final checkpoint. Its three seeds used benchmark windows with the official filter. METIS~\cite{metis2026} received EEG in volts. LEAF~\cite{leaf2025} used its released v1.0 checkpoint in three modes, with no instruction, the question alone, or the question and candidate prototypes.

For cognitive EEG analysis, CogEEGAgent maps natural-language questions to registered analysis workflows~\cite{hou2026cogeegagent}.
EEGJev uses questions to select candidate answers for supplied EEG windows.

\suppnote{Compute and parameters}
\label{app:compute}

The decision-head stage of EEGJev trains 806.8 million parameters, and the model with its frozen CBraMod encoder has
811.7 million (899.6 million with the 19 per-source encoders), to which the frozen MPNet text encoder (109.5 million)
is added. The 52 sets of spatial-token hidden layers account for 804,692,000 of the 806,770,979 trainable parameters
(99.74\%). Their first layer grows with the number of channels and the window length. The
hidden layers of the ordinary classifier are large for the same reason. Offline evaluation of EEGJev on one CPU thread at
32-bit precision took 6.8 to 7.6 hours per seed for the same 649,834 D1 and D2 test questions. Across the three runs,
the average was about 0.04\,s per question, including scoring and writing the results.

\paragraph{Single-question response time.}
We timed EEGJev and NeuroLM-B on the same 192 binary D1 validation questions about walking transitions (Exo)
and eye state (LEMON). For each source, we used 32 questions at each window length (1, 2 and 4\,s).
We tested both models on one Intel Xeon Gold 6126 CPU core (2.60\,GHz), then repeated the comparison on one
NVIDIA RTX PRO 6000 Blackwell Max-Q GPU. Both models used 32-bit floating point. TensorFloat-32 was disabled on the GPU.
On each platform, for each of three training seeds, the models answered 12 separate warm-up questions followed by the
192 timed questions one at a time. We alternated which model answered first. Timing began with the preprocessed EEG
window in host memory and the question and options available as text. It included question and option encoding and
scoring for EEGJev, and tokenization, EEG input assembly, generation of five tokens and answer parsing for NeuroLM.
For the GPU run, we included the input transfer and synchronized the device at both ends of each timed call.
Model loading, signal acquisition and offline preprocessing took place before timing.

For each model and platform, we pooled 576 timed calls. The 95th percentile was the shortest time by which at least
95\% of calls had returned an offered option. We assigned infinite time to a response that could not be parsed as an
offered option. On each platform, EEGJev returned an offered option on all 576 calls and NeuroLM on 554.
The NeuroLM-to-EEGJev P95 time ratios were 9.31 on the CPU and 2.60 on the GPU. Counting parsing failures at their observed return times
gave NeuroLM a 95th-percentile response time of 3,664.11\,ms on the CPU and 24.92\,ms on the GPU.

\suppnote{Shared and per-source encoders}
\label{app:persource}

On test D1, EEGNet is 0.7 points above the shared EEGJev
($-0.6$ to $+2.0$) and the shared ordinary classifier 0.3 above it ($-0.5$ to $+1.1$). On validation, the per-source
ordinary classifier is the highest model on D1 (66.9\%), and the per-source EEGJev is 1.2 points below EEGNet
($-2.1$ to $-0.2$). On D4, the per-source EEGJev is 1.3 points above the shared one ($+0.9$ to $+1.6$). On T2 it is
2.3 points below the shared one on test ($-3.1$ to $-1.5$) and not distinguishable on validation ($-0.1$, $-2.3$ to
$+2.5$). Under the held-out phrasings its seeds range from 53.7\% to 63.4\% (Table~\ref{tab:d1h}).
Table~\ref{tab:heldout} compares the two encoder settings on held-out tasks.

\suppnote{Held-out phrasings}
\label{app:d1h}

Table~\ref{tab:d1h} compares the test D1 questions with the same questions in the two held-out phrasings of each
attribute, which occur in no training text.

\begin{table*}[htbp]
\centering
\caption{Test D1 and D1h source-balanced mean balanced accuracy in \%, seed means, with the range over seeds for
D1h. EEGNet and LaBraM do not read the question and have no D1h row.}
\label{tab:d1h}
\EEGJevTableFont
\begin{tabular}{@{}lccc@{}}
\toprule
\EEGJevTableHead \textbf{Model} & \textbf{D1} & \textbf{D1h (seed range)} & \textbf{Change} \\
\midrule
\cellcolor{EEGJevTableTeal}EEGJev & 63.8 & 61.7 (59.7--62.9) & $-2.1$ \\
\cellcolor{EEGJevTableTeal}EEGJev, per-source & 64.1 & 59.8 (53.7--63.4) & $-4.3$ \\
Ordinary classifier & 64.2 & 64.2 (64.0--64.3) & $+0.0$ \\
Ordinary, per-source & 64.7 & 64.7 (64.2--65.0) & $+0.0$ \\
NeuroLM & 58.6 & 55.5 (51.0--58.3) & $-3.1$ \\
METIS & 43.5 & 42.8 & $-0.8$ \\
LEAF level 0 & 44.8 & 44.8 & $+0.0$ \\
LEAF level 1 & 45.0 & 44.9 & $-0.1$ \\
LEAF level 2 & 44.8 & 44.8 & $-0.0$ \\
\bottomrule
\end{tabular}
\end{table*}

\suppnote{Datasets no model has seen}
\label{app:unseen}

We also asked the frozen models left-or-right-hand motor imagery on two datasets that no model had been trained or
evaluated on, Weibo2014~\cite{yi2014weibo} with 10 persons and Dreyer2023A~\cite{dreyer2023large} with 60. The task
itself occurs in bciciv, so EEGJev, the ordinary classifier, EEGNet and LaBraM use their bciciv parts. The new
recordings are laid out on the 59 bciciv electrodes. Electrodes that a dataset has are copied, and each missing one is
the mean of its two nearest recorded neighbours on the official bciciv coordinates (18 missing in Weibo2014, 32 in
Dreyer2023A). LEAF maps the native electrodes with its own table. Windows start at the cue.

EEGJev stays at chance on both datasets (49.6\% to 52.4\%) and is below
the ordinary classifier in all six conditions, by 2.2 to 7.0 points with intervals that exclude zero. EEGNet reaches
52.5\% to 57.5\%. LEAF does best on Weibo2014 (63.6\% at level 2 and 2\,s) but is at chance on Dreyer2023A, and
LaBraM's bciciv models output almost a single class. No model does well. In the bciciv training data only two
persons imagine left and right hand movements, so this task has little training signal to begin with.

We tested whether a few labelled windows from a new person recover the task. For each person and
window length, $k$ labelled windows per class calibrate a logistic regression on the frozen features of the model,
and the person's other windows are scored, with 20 random draws of the labelled windows. With 20 windows per class,
EEGJev reaches 51.9\% to 60.8\% and the ordinary classifier 55.4\% to 69.9\%.
We average scores over participants, draws and three seeds.

\suppnote{Fixed options and text controls}
\label{app:controls}

\paragraph{Fixed EEG and options.}
We asked two questions about the same window, one for each of two attributes. Both offered the same four to ten
labels in the same order, with disjoint label sets for the two attributes. All tested pairs were available during
training. Attribute match required selecting a label of the requested attribute. Answer balanced accuracy required
selecting the correct label. Uniform choice uses the fraction of the offered labels that belong to the requested
attribute. Each split had 190 eligible combinations of source, window length, attribute pair and requested
attribute, with 409,461 validation questions and 380,570 test questions. The source-balanced means across three
seeds were 99.71\% attribute match and 63.25\% answer balanced accuracy on validation, and 99.70\% and 62.04\% on test
(Figure~3a).

For the paired score, both attributes had to meet the benchmark's eligibility criteria. The 90 eligible pairing
conditions yielded 201,963 validation pairs and 187,770 test pairs. A window could contribute to several attribute
pairs. Both answers were correct on 45.19\% of validation pairs and 44.37\% of test pairs after averaging within each
condition, within each source and then across sources.

\paragraph{Changes to questions and options.}
We scored the same D1 windows on validation and test with the frozen models of all three seeds. Scores were
similar when we re-ran the training phrasings. Table~\ref{tab:controls} reports each change, including the use of
an untrained task code such as “Task ID: T001”.

\begin{table*}[htbp]
\centering
\caption{Question and candidate controls of EEGJev on the 109 D1 tasks. BA is the source-balanced mean balanced
accuracy in \%, averaged over three seeds. Change is the paired difference to the formal model of the same seed in
points, with a 95\% interval from a bootstrap over tasks within sources. With letter codes, answers are scored by the
original classes.}
\label{tab:controls}
\setlength{\tabcolsep}{3pt}
\EEGJevTableFont
\begin{tabular}{@{}J{121pt}cccc@{}}
\toprule
 & \multicolumn{2}{c}{\cellcolor{EEGJevTableTeal}Validation} & \multicolumn{2}{c}{\cellcolor{EEGJevTableOrange}Test} \\
\cmidrule(lr){2-3}\cmidrule(l){4-5}
\EEGJevTableHead \textbf{Condition} & \textbf{D1 BA} & \textbf{Change} & \textbf{D1 BA} & \textbf{Change} \\
\midrule
\textbf{Formal model} & 64.4 & & 63.8 & \\
Training phrasings, scored again & 64.4 & $0.0$ ($-0.1$, $0.0$) & 63.9 & $0.0$ ($0.0$, $+0.1$) \\
Held-out phrasing 1 & 61.8 & $-2.6$ ($-3.3$, $-2.0$) & 60.9 & $-2.9$ ($-3.8$, $-2.0$) \\
Held-out phrasing 2 & 63.2 & $-1.2$ ($-1.6$, $-0.8$) & 62.6 & $-1.3$ ($-1.7$, $-0.9$) \\
No question & 62.6 & $-1.8$ ($-2.2$, $-1.4$) & 62.5 & $-1.3$ ($-1.7$, $-0.9$) \\
Candidates rotated & 64.4 & $0.0$ ($0.0$, $0.0$) & 63.8 & $0.0$ ($0.0$, $0.0$) \\
Question of another task & 51.3 & $-13.1$ ($-14.7$, $-11.5$) & 51.2 & $-12.6$ ($-14.1$, $-11.1$) \\
Untrained task code & 62.1 & $-2.4$ ($-2.8$, $-1.9$) & 62.0 & $-1.9$ ($-2.4$, $-1.3$) \\
Letter codes, meaning stated & 43.5 & $-21.0$ ($-22.9$, $-19.1$) & 43.5 & $-20.3$ ($-22.0$, $-18.5$) \\
Letter codes, meaning not stated & 43.6 & $-20.9$ ($-22.8$, $-19.0$) & 43.5 & $-20.3$ ($-22.1$, $-18.5$) \\
\bottomrule
\end{tabular}
\end{table*}

\suppnote{Inference rules for known tasks}
\label{app:knownrules}

Both rules use only saved outputs of the frozen models and apply to tasks whose full answer set was seen in training.
Both rules were fixed before v1.3 test evaluation and applied to the saved v1.3 validation and test outputs
without additional fitting.

\paragraph{Full-label check for D4.} For a D4 question, the model first scores the complete answer set of the task,
that is, the options of the matching D1 question of the same window. If the best of these answers is among the
offered options, the model answers with it, and otherwise it refuses. Across the same 31 test tasks, the full-label check
gives group-balanced scores of 65.4\% for EEGJev,
66.7\% for EEGNet and 65.5\% for the ordinary classifier. The paired difference (EEGJev minus EEGNet) is
$-1.3$ percentage points (95\% interval $-2.0$ to $-0.7$ from resampling tasks within sources). EEGJev's Act head scores 59.2\%.

\paragraph{Averaging with the ordinary classifier.} For D1 and D2 questions, the log-softmax over the offered options
of EEGJev and of the ordinary classifier of the same seed are averaged with weight 0.5 each. The weight was fixed at 0.5
before v1.3 test evaluation. Both models share the step-1 encoder, so part of the gain is the effect of combining
two output layers.

The averaged D2 answers had a joint score of 42.1\% on the test set.

\suppnote{D3 without the further-trained branch}
\label{app:d3}

The frozen decision head never saw D3 training questions. Table~\ref{tab:d3joint} compares its D3 answers
with those of the further-trained branch.

\begin{table*}[htbp]
\centering
\caption{D3 answers of the frozen decision head and the further-trained D3 branch. BA is balanced accuracy (\%).
Dist.\ is mean absolute bin distance.}
\label{tab:d3joint}
\setlength{\tabcolsep}{3pt}
\EEGJevTableFont
\begin{tabular}{@{}J{137pt}rccccc@{}}
\toprule
 & & \multicolumn{3}{c}{\cellcolor{EEGJevTableOrange}Test} & \multicolumn{2}{c}{\cellcolor{EEGJevTableTeal}Validation} \\
\cmidrule(lr){3-5}\cmidrule(l){6-7}
\EEGJevTableHead \textbf{Model} & \textbf{Tasks} & \textbf{BA} & \textbf{minus EEGNet} & \textbf{Dist.} & \textbf{BA} & \textbf{Dist.} \\
\midrule
EEGJev, decision head without the D3 branch & 22 & 35.3 & $-7.2$ ($-9.2$, $-5.1$) & 1.23 & 34.1 & 1.28 \\
EEGJev per-source, decision head without the D3 branch & 22 & 33.7 & $-8.7$ ($-10.4$, $-6.9$) & 1.20 & 34.3 & 1.21 \\
EEGJev, D3 branch & 22 & 41.5 & $-1.0$ ($-2.8$, $+0.8$) & \textbf{0.80} & \textbf{41.3} & \textbf{0.82} \\
EEGJev per-source, D3 branch & 22 & \textbf{41.6} & $-0.8$ ($-2.3$, $+0.6$) & 0.82 & 39.6 & 0.84 \\
\bottomrule
\end{tabular}
\end{table*}

\suppnote{External comparison with CBraMod}
\label{app:external}

EEGJev and CBraMod~\cite{wang2025cbramod} used the same splits from the released CBraMod preprocessing, with
three seeds and model-specific validation selection. On PhysioNet-MI~\cite{schalk2009eegmmidb,bci2000,pollard2026physionet},
their test balanced accuracies were 59.3\% and 61.7\%, respectively. On Mumtaz2016~\cite{mumtaz2016mdd,mumtaz2017wavelet},
they were 93.6\% and 89.3\%. The CBraMod paper reports 10 Mumtaz2016 test participants, while our split had 11.
EEGJev's gain came mainly from one participant with depression. The CBraMod reruns scored below the published
five-seed values of 64.2\% and 95.6\%. Mumtaz2016 reports ethics approval and written consent~\cite{mumtaz2017wavelet}.
The PhysioNet-MI dataset page does not report ethics approval.

\suppnote{Additional results and source data}
\label{app:figure-scores}

The paired D4 comparisons use common tasks and give 95\% intervals for the shared and per-source Act heads against
confidence rules and against each other. Scores are averaged over seeds within each task.

LEAF levels 0 and 1 use uncalibrated probabilities for D4 threshold fitting because their calibrated probabilities
are uniform. LaBraM, NeuroLM and LEAF cover 14 of the 22 D3 tasks. Fixed rules for per-source EEGJev are available
for T2 only.

The additional comparison with trained T2 readouts covered 3,307 test events, each placed once in each half of a window. On the 13 state-transition tasks, retaining the order of one-second probabilities improved the trained readouts by 1.86 percentage points for EEGNet and 0.99 points for the shared ordinary classifier. On the two tasks with a target in a stimulus stream, the gains were 14.29 and 18.51 points, respectively.

\noindent\begin{minipage}{\linewidth}
Supplementary Table~\ref{tab:heldout} gives the held-out task scores for Figure~3c,d. The text controls in Figure~3b are reported in Supplementary Table~\ref{tab:controls}.

\par\smallskip
\begingroup
\makeatletter
\def\@captype{table}
\makeatother
\centering
\caption{Held-out tasks, evaluated on the task's validation D1 rows (vrnf eye state at 1, 2 and 4\,s, asym face gender
at 1\,s). BA in \%, seed means. In training is the full model of the same encoder type on the same rows. The retrained
models ran on A100 slices. The ordinary classifier, EEGNet and LaBraM have no output for a task removed from
training.}
\label{tab:heldout}
\setlength{\tabcolsep}{2.5pt}
\EEGJevTableFont
\begin{tabular}{@{}J{102pt}J{71pt}cccc@{}}
\toprule
\EEGJevTableHead \textbf{Held-out task} & \textbf{Model} & \textbf{BA (seeds 0 / 1 / 2)} & \textbf{AUROC} & \shortstack{\textbf{Shuffled}\\\textbf{EEG}} & \shortstack{\textbf{In}\\\textbf{training}} \\
\midrule
\textbf{vrnf eye state (240 rows)} & \cellcolor{EEGJevTableTeal}EEGJev & 78.9 (80.8 / 76.7 / 79.2) & 0.86 & 48.6 & 83.1 \\
 & \cellcolor{EEGJevTableTeal}EEGJev, per-source & 81.2 (75.8 / 85.0 / 82.9) & 0.89 & 50.3 & 87.2 \\
\textbf{asym face gender (3,909 rows)} & \cellcolor{EEGJevTableTeal}EEGJev & 51.7 (60.3 / 48.4 / 46.5) & 0.53 & 49.7 & 82.1 \\
 & \cellcolor{EEGJevTableTeal}EEGJev, per-source & 47.9 (44.0 / 49.4 / 50.2) & 0.45 & 50.1 & 82.5 \\
\bottomrule
\end{tabular}
\par\endgroup
\end{minipage}

\FloatBarrier
\endgroup

\end{document}